\documentclass[prl,twocolumn,showpacs,groupedaddress,superscriptaddress]{revtex4}
\usepackage{graphicx}
\usepackage{amsmath,amssymb}

\begin{document}

\title{Interrelation between the distributions of kinetic energy release 
and emitted electron energy following a decay of electronic states}

\author{Ying-Chih Chiang}
\author{Frank Otto}
\author{Hans-Dieter Meyer}
\author{Lorenz S. Cederbaum}
\affiliation{Theoretische Chemie, Universit\"at Heidelberg,
    Im Neuenheimer Feld 229, D--69120 Heidelberg, Germany}

\date{\today}

\begin{abstract}
In an electronic decay process followed by fragmentation the kinetic energy release
and electron spectra can be measured. Classically they are the mirror image of each other,
a fact which is often used in practice. Quantum expressions are derived for both
spectra and analyzed. It is demonstrated that these spectra carry complementary
quantum information and are related to the nuclear dynamics in different 
participating electronic states. Illustrative examples show that
the classical picture of mirror image can break down and shed light on 
the underlying physics. 
\end{abstract}


\maketitle


The fragmentation of molecules and clusters following electronic excitation 
is a broad and widely studied process \cite{Eberhardt83,Baumert90,Aksela95}. The detailed knowledge of the 
process including the distribution of the produced fragments is relevant for 
atmospheric chemistry, for astrophysics and for basic research. 
The electronic excitation by photon or electron impact leads to excited electronic states 
which can dissociate by themselves or decay by emitting a photon or an electron, 
and the final states of this decay may then fragment. Photodissociation \cite{Schinke} provides 
a prominent class of examples of the former case, and the fragmentation of molecules 
which have undergone Auger decay \cite{Svensson89} forms an important class of examples of the latter. 

We shall concentrate here on the fragmentation following the decay of the 
excited electronic state and, to be specific, consider decays by emission of an electron, 
although, the situation is similar when a photon is emitted. 
Assume for instance a diatomic molecule AB where a core electron of A is removed by a high energy photon. 
The resulting $\text{A}^+ \text{B}$ will undergo Auger decay and emit an electron, the Auger electron. 
If dications $\text{A}^+ \text{B}^+$  are produced, they will, of course, be subject to a fragmentation 
into  $\text{A}^+$  and  $\text{B}^+$  via Coulomb explosion \cite{Eberhardt87,Hitchcock88}. 
Another interesting whole class of processes is interatomic Coulombic decay (ICD) 
\cite{Lenz97,Robin00,Marburger03,Jahnke04,Morishita06,Nico10_1,Jahnke10,Mucke10}.  
Assume an atom M and a rather distant neighboring atom N which is not chemically bound to M, 
an example would be a noble gas dimer. After removing a core electron from M, the Auger decay 
is essentially atomic and $\text{M}^{2+}$ is produced. ICD then takes place as a follow up of 
Auger resulting in the $\text{M}^{2+} - \text{N}^+$  triple ion and an emitted electron, the ICD electron 
\cite{Robin03,Morishita06,Liu07,Morishita08,Kreidi08_1,Kreidi08_2,Yamazaki08,Philipp09}. 
The ion, of course, undergoes Coulomb explosion and fragments.

\begin{figure}
\centering 
\includegraphics[width=7.5cm,height=5.5cm]{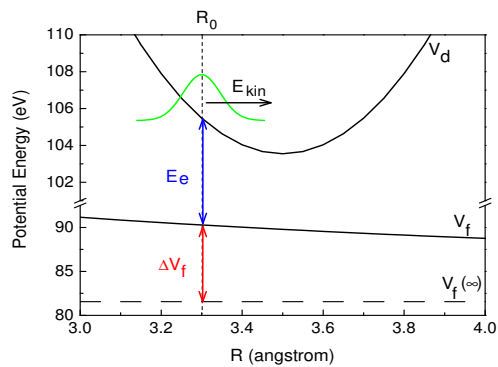} 
\vspace{-0.5cm}

\caption{(Color online) Potential energy curves of a system
which undergoes an electronic decay process with subsequent 
fragmentation. In a classical picture the system decaying at time $t$ at an
internuclear distance $R_0$ will produce emitted electrons with kinetic 
energy $E_e(R_0)$ and the fragments with KER energy $E_{\text{KER}}(R_0) = E_T - E_e(R_0)$. 
Two sources contribute to the KER: the energy released from the potential 
difference $\Delta V_f=V_f(R_0)-V_f(\infty)$, and the kinetic energy $E_{\text{kin}}$.
In the quantum picture the wavepacket $\psi_d$ propagates 
and continuously decays onto the final potential curve $V_f$ where the 
resulting wavepacket $\psi_f$ propagates. The KER and electron spectra are 
then given by Eqs.~(\ref{eq:ker2},\ref{eq:espect2}).
}
\label{fig:potential}
\end{figure}

In such experiments one measures the distributions of the total kinetic energy released 
by the fragments (KER spectrum) 
and the kinetic energy of the electron emitted 
in the decay (electron spectrum) which can be obtained separately or from 
the coincidence measurements of all charged particles \cite{Jahnke04,Morishita06,Liu07,Morishita08,Kreidi08_1,Kreidi08_2,Yamazaki08,Jahnke10}. 
The question immediately arises whether at all the KER and electron spectra contain 
complementary information on the system and the underlying process. 
Let us start discussing the problem in the framework of classical mechanics. 
Let the excited molecule possess the total energy $E_{T'}$ which consists of the kinetic energy  
$E_{\text{kin}}$ of the relative nuclear motion and the potential energy $V_d$. 
At a specific time $t$ the excited molecule decays at an internuclear distance $R_0$ 
and emits an electron of energy $E_e(R_0)$ which is given by the energy difference $V_d(R_0) - V_f (R_0)$ 
between the potential energy curves $V_d$ of the decaying and $V_f$ of the final electronic states. 
The situation is depicted in Fig.~\ref{fig:potential}. The produced ion now moves on the dissociative curve $V_f$ 
and fragments. The KER energy $E_{\text{KER}}$ is obviously the difference between 
the energies of the ion produced at $R_0$ and that of the fragments after they have 
completely separated: $E_{\text{KER}} = [E_{\text{kin}} +V_f (R_0)] - V_f (\infty)$.  
The consequences are clear. If we know the potential energy curves $V_d(R)$  and $V_f(R)$ 
of the decaying and final electronic states, there is nothing new in the KER spectrum 
$\sigma_{\text{KER}}(E_{\text{KER}})$ compared to the electron spectrum $\sigma_e(E_e)$ and vice versa. 
Because of energy conservation at every point $R$, one spectrum can be expressed as the mirror image of the other:  
$\sigma_{\text{KER}}(E_{\text{KER}}) = \sigma_e(E_T - E_e)$ , where $E_T$ is the total 
energy available relative to the total electronic energy of the separated atomic fragments, 
i.e., $E_T =  E_{\text{KER}} + E_e = E_{T'} - V_f (\infty)$.    

Indeed, the mirror image procedure has been successfully verified
in experiments and also computationally, see e.g. \cite{Jahnke04, Simona04} .  
Are there no fingerprints of quantum dynamics which make the spectra inequivalent? 
We shall show below that quantum dynamics can easily lead to a different physical content 
of the two spectra and then the mirror image procedure breaks down. 

We proceed by deriving quantum expressions for the KER and electron spectra 
starting from the coincidence spectrum $\sigma(E_{\text{KER}} ,E_e,t)$ where all particles 
are measured in coincidence as a function of time. 
The KER spectrum is determined from the coincidence spectrum by integrating over all electrons 
\begin{equation}
\sigma_{\text{KER}}(E_{\text{KER}},t) = \int \text{d} E_e \,\,\, \sigma(E_{\text{KER}},E_e,t).
\label{eq:ker1}
\end{equation}
and the electron spectrum by integrating over all $E_{\text{KER}}$
\begin{equation}
\sigma_{e}(E_e,t) = \int \text{d} E_{\text{KER}}  \,\,\, \sigma(E_{\text{KER}},E_e,t).
\label{eq:espect1}
\end{equation}
In current experiments the spectra are measured after a long time, i.e., $t \to \infty$, 
and we will in our explicit examples concentrate on this standard case. 
We would like to point out, however, that time-dependent spectra carry much more 
information on the physics of the underlying processes, see e.g. \cite{Chiang10}. 

One can further evaluate Eqs.~(\ref{eq:ker1},\ref{eq:espect1}) by 
noting that the coincidence spectrum is determined by the nuclear wavepacket 
$\psi_f(E_e,t)$ propagating on the potential energy curve $V_f(R)$ of the final 
electronic state to which the system has decayed \cite{Elke96}. 
Before the decay has started this wavepacket is, of course, unpopulated, 
i.e., its value is zero. As time proceeds, $\psi_f(E_e,t)$ becomes populated and 
carries all the information of the decayed system. Its equation of motion follows 
from the Schr\"odinger equation and is well known \cite{Elke96} :
\begin{equation}
i \vert \dot{\psi}_f(E_e,t)\rangle = W_{d \to f}|\psi_{d}(t)\rangle + (\hat{H}_{f} +E_e)|\psi_f(E_e,t)\rangle .
\label{eq:mastereqf}
\end{equation}
$\hat{H}_f$  is the common nuclear Hamiltonian governing the nuclear motion 
on the potential curve $V_f(R)$, and the transition matrix element from the decaying state 
to the final state is denoted by $W_{d \to f}$ . The initial excitation of the decaying state is, 
of course, provided by experiment. Once excited, the nuclear wavepacket $\psi_d(t)$ of this electronic state 
propagates on the complex potential curve $V_d - i \Gamma/2$, where $\Gamma$ is the total decay rate 
of the state and may depend on $R$. With time electrons are emitted and $\psi_d(t)$ decays 
and loses its norm thus populating $\psi_f(E_e,t)$ as can be seen in Eq.~(\ref{eq:mastereqf}),  
in which $W_{d \to f} \psi_d(t)$ is the source term for populating the final state. 
There is an intimate relation between $\Gamma$ and $W_{d \to f}$. 
If there are several final electronic states $f$ into which the system decays, 
then $\Gamma_f  = 2 \pi | W_{d \to f} |^2$  is the partial decay rate and 
the total rate is simply $\Gamma = \sum_f  \Gamma_f$. 
In the following we assume for simplicity of presentation a single final state. 
We note, however, that the extension to several final states is trivial. 

The coefficients  in the expansion of $\psi_f(E_e,t)$ in the complete set of 
dissociating nuclear wavefunctions $|E_f \rangle$ of the potential $V_f(R)$ determine 
as usual the probability of finding the fragments with energy $E_f$ and the 
electron with energy $E_e$ and thus determine the coincidence spectrum \cite{Nico10_2}. 
Explicitly, $| \psi_f(E_e,t) \rangle  = \sum c_{E_f} (E_e,t)  |E_f \rangle$, where $|E_f \rangle$ is an eigenfunction of 
$\hat{H}_f$ with energy $E_f = E_{\text{KER}} + V_f (\infty)$,  immediately leads to 
$\sigma (E_{\text{KER}},E_e,t) = |c_{E_f} (E_e,t)|^2$.   
Inserting this expansion into Eq.~(\ref{eq:mastereqf}) and projecting on a single energy 
allows one to find the solution of the expansion coefficient for that energy. 
Integrating over the electron energy $E_e$ gives via Eq.~(\ref{eq:ker1}) 
an interesting expression for the KER spectrum:
\begin{eqnarray}
\label{eq:ker2}
\sigma_{\text{KER}}(E_{\text{KER}},t)=2 \pi \int^{t} \text{d} t' 
| \langle  E_f |W_{d \to f} | \psi_{d} (t') \rangle |^2 \:.
\end{eqnarray}

This general expression is easily interpreted. First, the KER spectrum is solely 
determined by the dynamics in the decaying state and not at all by the dynamics 
in the final state. Second, as in any transition, the integrand can be viewed as 
a generalized Franck-Condon factor connecting $\psi_d$ and the dissociative eigenfunction 
of the final state. Third, and importantly, since $\psi_d$ is time dependent, the KER spectrum 
at time $t$ is given by the Franck-Condon factor which has accumulated up to this time, 
or briefly, by the \textit{accumulated Franck-Condon factor}.

Similarly, we obtain for the electron spectrum 
\begin{equation}
\label{eq:espect2}
\sigma_e (E_e, t) =\langle \psi_f(E_e,t) | \psi_f(E_e,t) \rangle,  
\end{equation}
which is, however, a well-known result \cite{Elke96,Lenz93}. In sharp contrast to the KER spectrum, 
the electron spectrum is determined by the norm of the final wavepacket which depends on 
the energy  of the emitted electron, and thus solely reflects the dynamics in the final state. 
Of course, this dynamics itself is not independent of that in the decaying state as can be seen 
in Eq.~(\ref{eq:mastereqf}). Accordingly, $\psi_d(t)$ evolves and decays and its losses 
$W_{d \to f} \psi_d(t)$ are continuously transferred to the final state where they continue 
to propagate but now with the final state Hamiltonian $\hat{H}_f$ . These transferred losses can, 
in principle, collide with the already available parts of $\psi_f(t)$ and interfere with them. 
All of these rather complex dynamical phenomena are missing in the accumulated 
Franck-Condon factor which determines the KER spectrum. In this respect the electron spectrum 
carries much more intricate information. We would like to remind, however, that the differences 
between the two kinds of spectra are due to quantum effects.

After having derived and interpreted the full quantum expressions for 
the KER and electron spectra we present illustrative examples. A transparent model
example and two realistic examples which have been recently measured.
In the model example depicted in Fig.~\ref{fig:potential}, the decaying state is described by
a harmonic potential curve $V_d$ (vibrational frequency 200 meV) and
the final state of the decay by the dissociative curve $2/R+V_f(\infty)$ which is chosen
to be the same as in one of the realistic examples discussed below.
The decaying electronic state decays by a constant rate $\Gamma = 200 \,\,\, \text{meV}$
which is within the typical range of Auger decay rates \cite{Krause79,Ohno95}.
The system is initially in its ground electronic (and vibrational) state which is
also chosen to have a harmonic potential curve but which can be shifted 
with respect to $V_d$. As often done, the system is excited
by a broad band light pulse transferring its initial vibrational Gaussian
wavefunction $\psi_d(0)$ vertically to $V_d$.
This wavepacket $\psi_d(t)$ can swing to and fro on $V_d$
and thereby decay as electrons are continuously emitted.
The parts of $\psi_d(t)$ lost by the decay 
($d \langle \psi_d(t) | \psi_d(t) \rangle /dt = -\Gamma \langle \psi_d(t) | \psi_d(t) \rangle$)
appear as $\psi_f(E_e,t)$ on the final dissociative curve, see Eq.~(\ref{eq:mastereqf})
describing the dissociation dynamics of the system.

\begin{figure}
\centering 
\includegraphics[width=7cm,height=5cm]{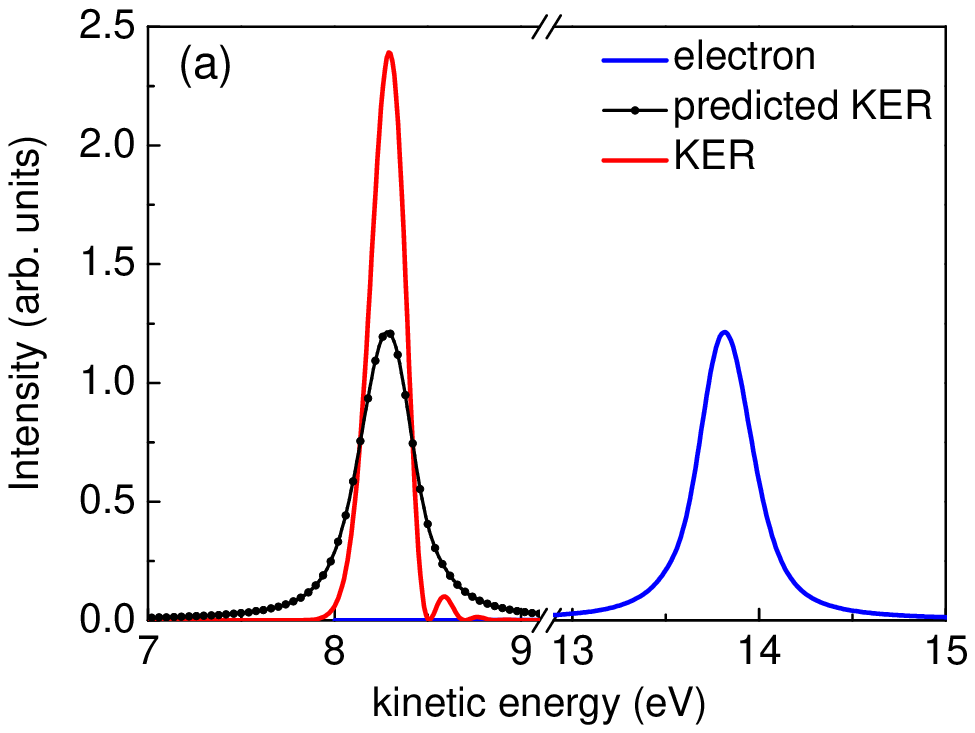}
\vspace{-0.5cm}

\includegraphics[width=7cm,height=5cm]{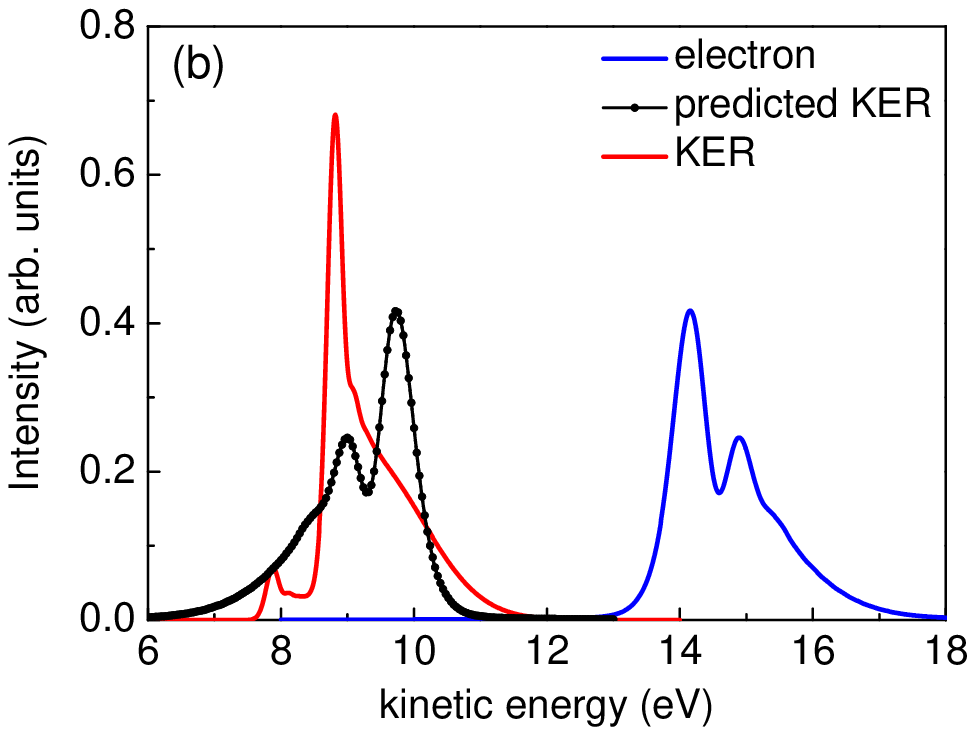} 
\vspace{-0.5cm}

\caption{ (Color online) The electron and KER spectra of a
model study (see text and Fig. 1). The initial wavepacket is chosen to 
be a Gaussian (lowest vibrational level of the ground electronic state of the system) 
in both panels: in panel (a) it centers at 3.5 \AA, in panel (b) at 3.3 \AA. 
$\Gamma$ and the potential curves are given in the text. 
These spectra are computed via Eqs.~(\ref{eq:ker2},\ref{eq:espect2})
in the limit $t \to \infty$. The results are compared with the (classically) 
predicted KER spectrum obtained by mirroring the electron spectrum. The differences between the 
predicted and exact KER spectra (shown at the left-hand side in both panels 
by the black dotted and red lines, respectively) demonstrate that quantum effects can be substantial.}
\label{fig:fig2}
\end{figure}
The results of the numerical calculations on the model example are collected in 
Fig.~\ref{fig:fig2}. Let us begin by choosing equilibrium distances of the ground and 
decaying states to be the same. Then, only the lowest vibrational level of $V_d$
is populated by the optical transition. Correspondingly, no nuclear dynamics
takes place on $V_d$, and one has $\psi_d(t)=\psi_d(0)e^{-\Gamma t/2}$.
What to expect in such a simple situation?
Eq.~(\ref{eq:ker2}) can be easily solved giving 
$\sigma_{\text{KER}}(E_{\text{KER}},\infty)=|\langle E_{\text{KER}}+V_f(\infty)|\psi_d(0)\rangle|^2$,
i.e., the KER spectrum is determined by the usual Franck-Condon factors. The decay rate $\Gamma$
does not influence at all the KER spectrum! 
On the other hand, even in this simple situation,
this is not the case for the electron spectrum, which turns out 
to be given by convoluting the mirror image of $\sigma_{\text{KER}}(E_{\text{KER}},\infty)$ with a Lorentzian
of full width at half maximum (FWHM) $\Gamma$.
The numerical results are shown in Fig.~\ref{fig:fig2}a together with the KER spectrum
obtained as the mirror image of the electron spectrum as expected classically (denoted
`predicted KER').  
Clearly, the exact KER spectrum is narrower and exhibits a nodal structure
at its high-energy wing being a fingerprint of the dissociative eigenfunctions of state $f$.
According to the above, convoluting this KER spectrum with a Lorentzian of 
FWHM $\Gamma$ reproduces exactly the classically predicted KER.

Shifting the equilibrium distance of the system to smaller values ( 3.3 $ \text{\AA}$  in Fig.~\ref{fig:potential}),
introduces nuclear dynamics of $\psi_d$ on the decaying curve $V_d$. The decay process becomes more 
intricate even for a constant $\Gamma$ and harmonic curves, and the electron spectrum
exhibits vibrational interference effects which have been discussed theoretically and
measured \cite{Neeb94}. The computed electron spectrum in Fig.~\ref{fig:fig2}b indeed
shows a typical vibrational progression including the impact of the interference
effects. Interestingly, the exact KER spectrum possesses a very different 
structure and does not at all resemble the classically predicted picture of
mirror imaging the electron spectrum. Instead of a progression of declining peaks
ending with a long low-energy tail, the exact KER spectrum essentially consists of a 
small and a pronounced sharp peak at low energy and a long high-energy tail.
Different from the electron spectrum, where the structure shows the fingerprint 
of different vibrational levels of $V_d$, the sharp peak together with the small shoulder
of the KER spectrum are contributed by $\psi_d$ at the classical turning points.
This is verified by evaluating the time evolution of Eq.~(\ref{eq:ker2}) with
a semiclassical approximation for $|E_f\rangle$.

\begin{figure}
\centering 
\hspace{-0.4cm}
\includegraphics[width=4.5cm,height=4cm]{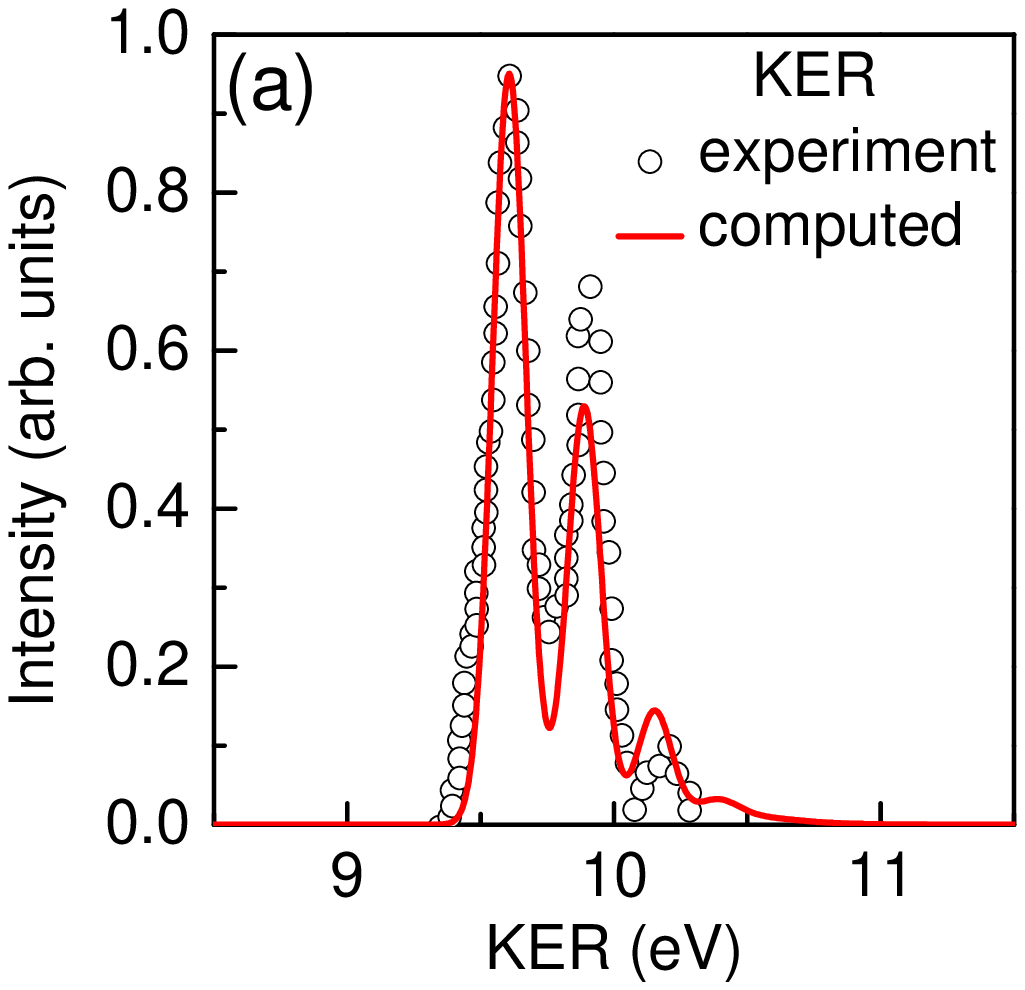} 
\hspace{-0.5cm}
\includegraphics[width=4.5cm,height=4cm]{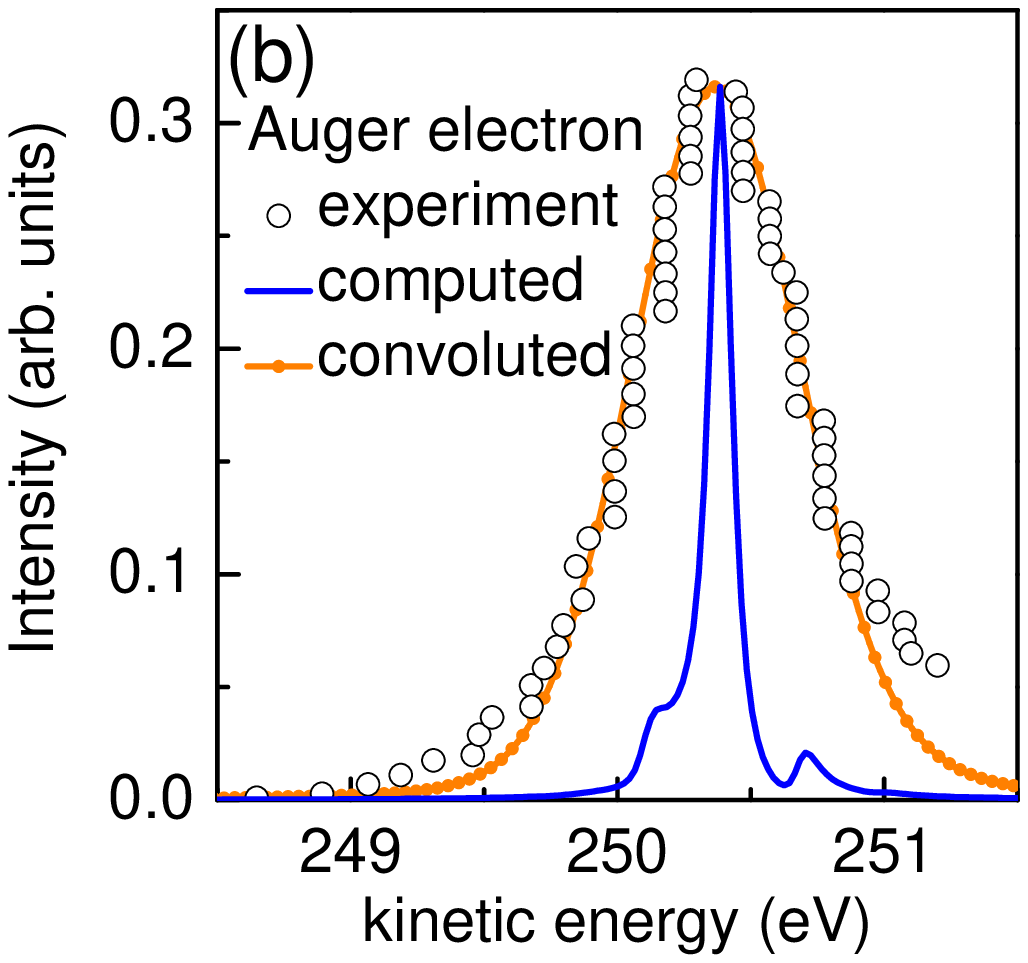} 
\vspace{-0.5cm}

\hspace{-0.4cm}
\includegraphics[width=4.5cm,height=4cm]{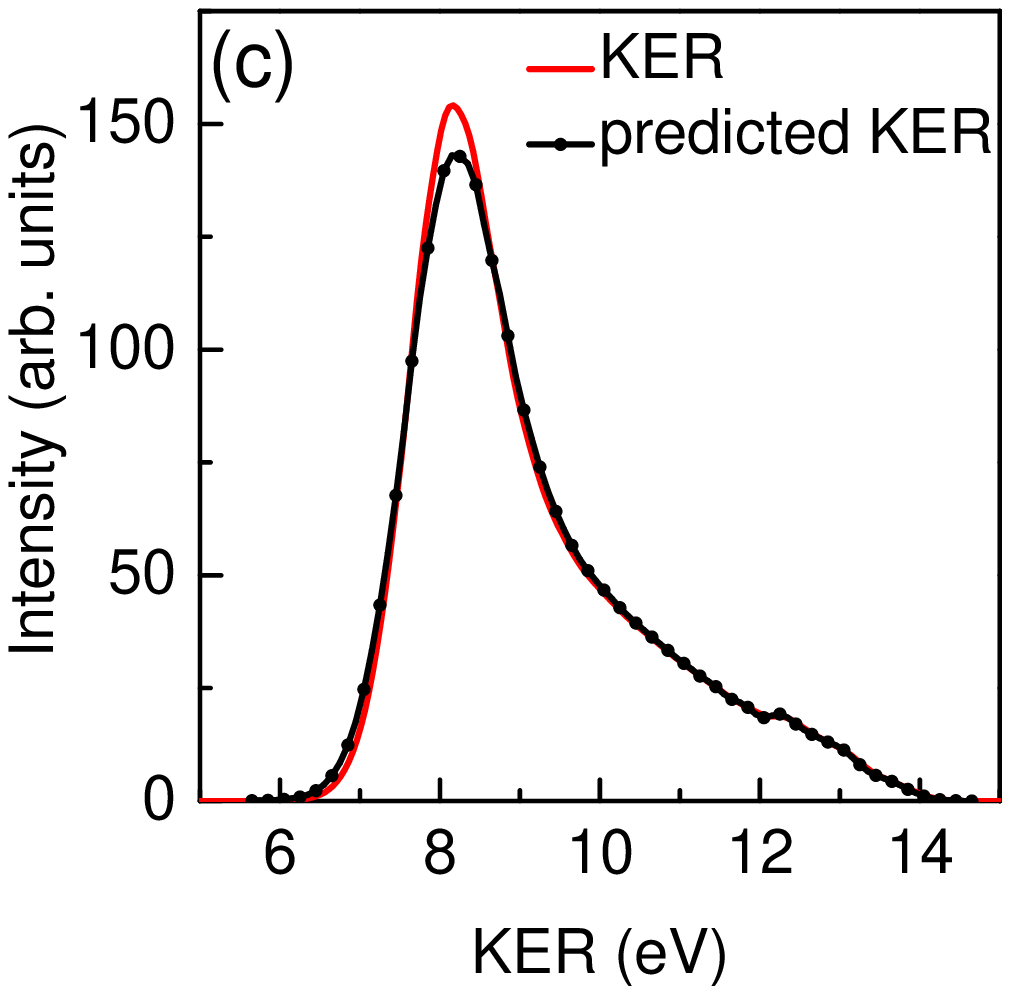}
\hspace{-0.5cm}
\includegraphics[width=4.5cm,height=4cm]{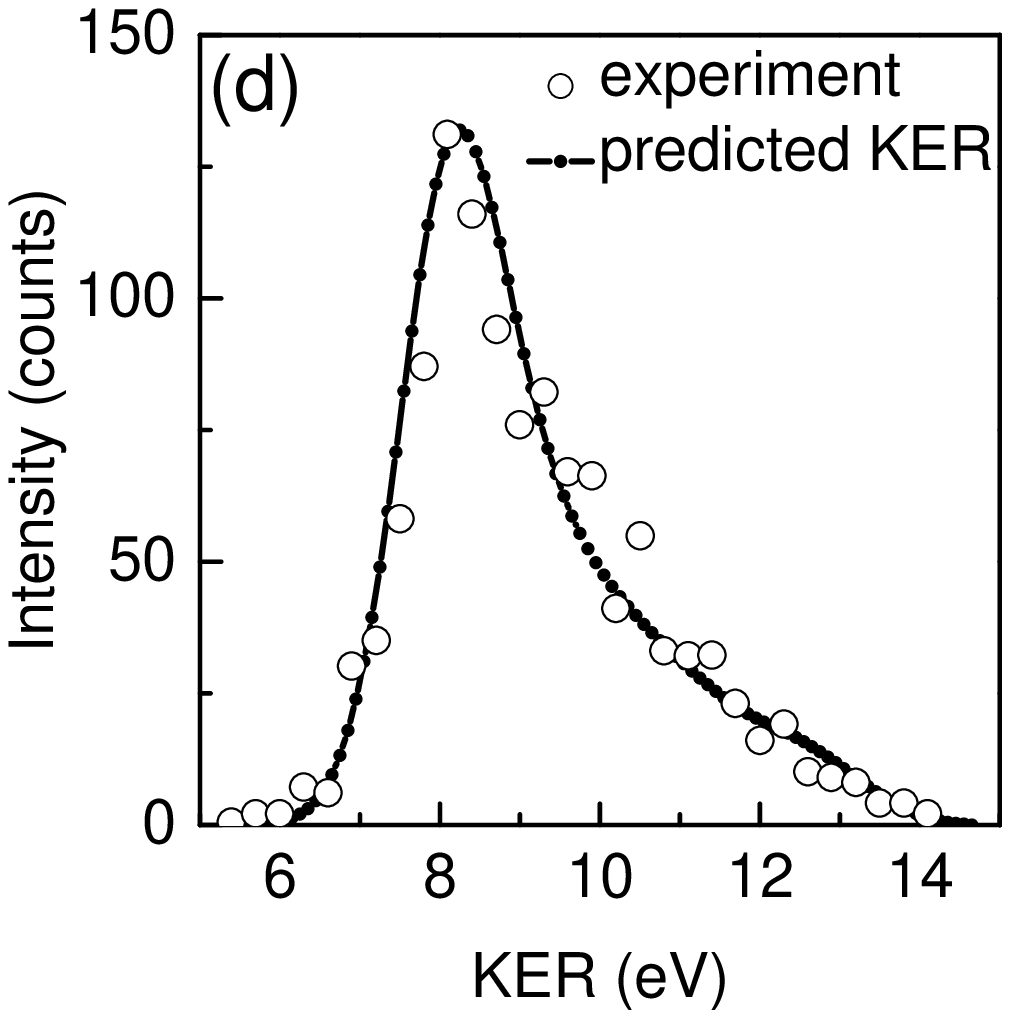}  
\vspace{-0.5cm}
\caption{ (Color online) 
Computed and experimental KER and electron spectra for two different processes. 
Upper panels: Auger process in CO ($\text{CO}^+ \to \text{CO}^{2+}(2\, ^1\Sigma^+) +e^-$). 
Lower panels: ICD in NeAr following the Auger decay of $\text{Ne}^+ \,\text{1s}^{-1}$ 
($\text{Ne}^{2+}(\text{2s}^{-1}\text{2p}^{-1}\, ^1\text{P})\text{Ar}
\to \text{Ne}^{2+}(\text{2p}^{-2}\, ^1\text{D})-\text{Ar}^+(\text{3p}^{-1})+e^-$). 
(a) Auger KER spectra. The experimental spectrum is from Ref. \cite{Weber01}.   
To compare with experiment, the computed KER spectrum (red curve) is convoluted 
with a Gaussian of 0.15 eV FWHM.  
(b) Auger electron spectra. The experimental spectrum is from Ref. \cite{Weber03}. 
Shown are also two computed spectra. The spectrum as calculated using Eq.~\ref{eq:espect2} (blue curve) 
and that obtained by convoluting it with a Gaussian of 0.68 eV FWHM to account for 
the experimental resolution (orange curve). 
Clearly, the electron and KER spectra are far from being the mirror image of each other. 
(c) Computed ICD spectra. Shown are the KER spectrum calculated via Eq.~\ref{eq:ker2} (red curve) 
and the ICD electron spectrum computed using Eq.~\ref{eq:espect2} and mirror imaged to provide 
a KER spectrum as predicted by the classical theory (black curve).    
(d) The experimental KER spectrum \cite{Ouchi11} is compared with the computed KER spectrum 
obtained by convoluting the black curve in panel (c) with a Gaussian of 0.7 eV FWHM \cite{Ouchi11} to account for 
the experimental resolution (black curve). Theory and experiment compare well. In this example the 
prediction of classical theory holds that the electron and KER spectra are the mirror image of each other 
(see also text). }
\label{fig:fig3}
\end{figure}
Let us now examine realistic examples, namely
the Auger decay of CO and ICD of NeAr.
Consider the Auger channel
$\text{CO}^+ (\text{C}\, \text{1s}^{-1}) \to \text{CO}^{2+}(2\,^1\Sigma^+) +e^-$,
whose electron and KER spectra have been measured \cite{Weber01,Weber03}.
Potential curves and $\Gamma$ are taken from literature \cite{Carroll02,Eland04}.
The experimental and computed KER spectra depicted in Fig.~\ref{fig:fig3}(a) coincide well.
The computed spectrum is computed via Eq.~(\ref{eq:ker2}) and broadened to account for the experimental resolution. This procedure is in accord with the discussion of the fragmentation process in \cite{Weber01,Weber03}.
The experimental and computed Auger electron spectra are shown in Fig.~\ref{fig:fig3}(b).  
Because of the difficulty to measure this spectrum, the resolution is much lower than for the KER spectrum \cite{Weber01,Weber03} 
and we show both the spectra computed via Eq.~(\ref{eq:espect2}) (blue curve) 
and the one convoluted with a Gaussian resolution function (orange curve). 
The latter reproduces the experiment. 
Regardless of the experimental resolution, it is obvious from these calculations that the KER and Auger electron spectra 
are far from being the mirror image of each other as expected classically. 
 
Contrary to the molecular Auger process, the electron and KER spectra
obtained for the ICD processes in noble gas dimers
are usually considered to be the mirror images of each other.
Here we study as an example of current interest the ICD process
$\text{Ne}^{2+}(\text{2s}^{-1}\text{2p}^{-1}\, ^1\text{P})\text{Ar}
\to \text{Ne}^{2+}(\text{2p}^{-2}\, ^1\text{D})-\text{Ar}^+(\text{3p}^{-1}) +e^-$
following the Auger decay of $\text{Ne}^+ (\text{1s}^{-1})$.
This very fast Auger decay produces $\text{Ne}^{2+}$ atomic dicationic states and some of them can 
further decay by ICD producing $\text{Ne}^{2+} \text{Ar}^+$ triply ionized states
which undergo Coulomb explosion.  
The process has recently been measured \cite{Ouchi11}.
In our calculations we utilize the potential curves of \cite{Philipp09} and the value of $\Gamma$ suggested in \cite{Ouchi11}. 
The mirror image of the computed ICD electron spectrum and the computed KER spectrum
are shown in Fig.~\ref{fig:fig3}(c) and agree well with each other.
This agreement is due to the facts
that the potential $V_d$ is very shallow and the rate $\Gamma$ is rather small. 
Consequently, several quasi-degenerate vibrational levels of $V_d$ are initially populated,
and the system behaves as if a single effective level is populated.
A comparison with experiment is provided in Fig.~\ref{fig:fig3}(d). 

Although being the mirror image of each other within the classical picture,
the KER and electron spectra of a decaying state carry complementary quantum information on 
the decay process. While the latter reflects the nuclear dynamics in the 
final electronic state and is sensitive to interference effects, the former
measures the accumulated Franck-Condon factor of the decay and
is the projection of the dynamics in the decaying electronic state
on the potential curve of the final state. The explicit general expressions derived allow
to compute and analyze these spectra, especially if a pulse is involved. Illustrative examples show that the classical 
picture of mirror image can break down due to quantum effects.

Y.C. C. would like to thank IMPRS-QD for financial support and 
Dr. Alexander Kuleff for helpful discussions. L.S.C. acknowledges
financial support by the DFG.

\end{document}